\documentclass[]{spie}  

\usepackage{amsmath,amsfonts,amssymb}
\usepackage{graphicx}
\usepackage[colorlinks=true, allcolors=blue]{hyperref}
\usepackage{xspace}
\usepackage{upgreek}

\newcommand{\umux}{$\upmu$mux\xspace}

\title{Phase Drift Monitoring for Tone Tracking Readout of Superconducting Microwave Resonators}

\author[a,b]{Max Silva-Feaver}
\author[c,d]{Zeeshan Ahmed}
\author[a,b]{Kam Arnold}
\author[c]{Josef C. Frisch}
\author[e]{John Groh}
\author[c,d]{Shawn W. Henderson}
\author[c]{Jesus Vasquez}
\author[c,f]{Cyndia Yu}
\affil[a]{University of California San Diego Department of Physics, 9500 Gilman Dr, La Jolla, CA, USA}
\affil[b]{Center for Astrophysics and Space Sciences, 9500 Gilman Dr, La Jolla, CA, USA}
\affil[c]{Kavli Institute for Particle Astrophysics and Cosmology, Menlo Park, CA, USA}
\affil[d]{SLAC National Accelerator Laboratory, 2575 Sand Hill Rd, Menlo Park, CA, USA}
\affil[e]{National Institute of Standards and Technology, 325 Broadway, Boulder, CO, USA}
\affil[f]{Stanford University Department of Physics, 450 Serra Mall, Stanford, CA, USA}

\authorinfo{Further author information: (Send correspondence to M.S.F.)\\M.S.F.: E-mail: msilvafe@ucsd.edu}

\begin{document} 
\maketitle

\begin{abstract}
A number of modern millimeter, sub-millimeter, and far-infrared detectors are read out using superconducting microwave (1-10GHz) resonators. The main detector technologies are Transition Edge Sensors, read out using Microwave SQUID Multiplexers (\umux) and Microwave Kinetic Inductance Detectors. In these readout schemes, sky signal is encoded as resonance frequency changes. One way to interrogate these superconducting resonators is to calibrate the probe tone phase such that any sky signal induced frequency shifts from the resonators show up primarily as voltage changes in only one of the two quadratures of the interrogation tone. However, temperature variations in the operating environment produce phase drifts that degrade the phase calibration and can source low frequency noise in the final detector time ordered data if left to drift too far from optimal calibration. We present a method for active software monitoring of the time delay through the system which could be used to feedback on the resonator probe tone calibration angle or to apply an offline cleaning. We implement and demonstrate this monitoring method using the SLAC Microresonator RF Electronics on a 65 channel \umux chip from NIST.
\end{abstract}


\section{INTRODUCTION} \label{sec:intro}
Superconducting microwave resonators fabricated in the $\mathcal{O}$(1-10 GHz) range are a rapidly growing technology for facilitating the readout of large format superconducting detector arrays for applications in low background X-ray\cite{Lynx_2019, CECIL_2012}, mm/sub-mm\cite{McCarrick_2021, Duell_2020}, and optical/infrared\cite{Nakada2020, Walter_2020} astronomy. In particular, both Microwave Kinetic Inductance Detectors (MKIDs)\cite{Day2003} and Microwave Superconducting Quantum Interference Device (SQUID) Multiplexer (\umux)\cite{mates_2011} readout architectures take advantage of the large readout channel counts and compact modules enabled by high density nanofabrication of superconducting resonators. In both architectures astrophysical signals induce shifts in the resonance frequencies of the resonators.

One way to read out these resonators is by generating a comb of probe tones each tuned to an individual resonator's frequency with a Digital to Analog Converter (DAC), transmitting the comb through the cryogenic resonators, and measuring the received comb on an Analog to Digital Converter (ADC). We detect changes in each resonator's frequency by measuring changes in the real and imaginary components of the forward transmission at the probe tone frequency. 

The SLAC Microresonator Radio Frequency (SMuRF) Readout electronics\cite{Henderson_2018} used for measurements in this article apply a calibration during the initial resonator ``tuning" procedure converting measured voltage on the ADC at each probe to an estimate of the frequency shift. This estimated frequency shift, called the frequency error ($df$), is then used to adjust the frequency of the tones such that $df$ is always kept at 0. Regularly updating the frequency of the probe tones allow SMuRF to tone-track keeping the probe tones on the resonance frequency which confers numerous benefits. Additionally, for \umux readout a flux ramp (typically a saw tooth ramp) is applied to the SQUIDs\cite{Mates2012} which transduces the detector signal to a phase shift of a periodic SQUID modulation. The SMuRF firmware adaptively demodulates the flux ramp modulation for all probe tones in real time and the phase of the demodulated signal contains the detector signal. 

Drifts in the round trip system time delay, however, spoil the calibration established during resonator tuning and produce a leakage into the detector time ordered data. These drifts are primarily caused by temperature-dependent length contraction in the coaxial cabling between the readout electronics and the cryostat. This effect can be corrected by monitoring the variation in system time delay and applying a correction to the channel-dependent angle rotation. For experiments that integrate for long periods of time and care about the low frequency noise profile such as Cosmic Microwave Background telescopes, estimation, monitoring, and cleaning of this leakage as we demonstrate in this proceedings is key to achieving stable noise performance over long observations. 

In this article we present a demonstration of phase drift monitoring and cleaning. In section \ref{sec:setup} we describe the experimental setup including the hardware configuration in section \ref{sec:hw_setup} and the resonator tuning procedure in section \ref{sec:tuning}. Section \ref{sec:res_tf} presents a detailed overview of the resonator transfer function and how it is affected by time delays, and section \ref{sec:expected_mag} estimates the level of drift we expect given some reasonable assumptions about the hardware design and operating environment. Section \ref{sec:measurement} is a measurement demonstration validating our assumptions about temperature coupling in section \ref{sec:expected_mag}. In section \ref{sec:phase_injection} we introduce a modified measurement setup designed to test phase drift cleaning and present the results of data cleaning method in section \ref{sec:results}. Lastly in section \ref{sec:conclusion} we conclude and provide some avenues for future work on this effort including implementation of an active feedback method.

\section{Measurement Setup}\label{sec:setup}
Here we describe the measurement setup and standard operating procedure to set up \umux resonators for readout of detectors using the SMuRF electronics. In section \ref{sec:hw_setup} we discuss the physical hardware connections in the readout chain and in section \ref{sec:tuning} we review the software and firmware steps used to set up the readout.
\subsection{Hardware Configuration}\label{sec:hw_setup}
Our measurement setup is at SLAC National Accelerator Laboratory in a large highbay-style assembly hall. All measurements shown in this paper were taken between June and July of 2022 during the North American summer time. 
The SMuRF electronics system used to generate the comb of RF probe tones, DC amplifier biases, and flux ramp signal is located in a standard computing rack. Next to the rack is a $\sim$4m tall vibration-isolating frame holding a Bluefors LD250 dilution refrigerator (DR). The warm RF coaxial cables used are Minicircuits CBL-Xm-SMSM+ (where X is the cable length), which run $\sim$3.5m from the SMuRF RF input/outputs to an interface plate that holds low noise amplifiers (LNAs) and fixed attenuators, then an additional $\sim$ 1m to the vacuum SMA feedthroughs on the top flange of the DR for a total of $\sim$9m of round-trip warm cabling. Next to the electronics rack is a water chiller used to cool the Cryomech PT415 compressor unit which turns on and off on a $\sim$10 minute cycle. When turned on it exhausts hot air on the electronics rack and warm cables, creating a $\sim$0.4$^{\circ}$C temperature increase in the local operating environment.

Inside of the DR unit is a series of low loss, low thermal conductivity coaxial cables and fixed attenuators that carry the signals from the RF input to the mixing chamber which we PID to a temperature of 100mK. The \umux chips used are the NIST umux100k\_v3.2 chips optimized for CMB applications\cite{Dober2020} installed into a single chip sample box just large enough for a 64 resonator chip, 2 RF coax-to-CPW transition RF Rogers circuit board, and a small DC FR4 DC interface board to carry in the flux ramp lines. On the output are two stages of cryogenic low noise amplification (at 4 Kelvin, and 40 Kelvin). The cryogenic RF components are similar to the design implemented for the Simons Observatory\cite{SO_RFChain}.

\begin{figure}
    \centering
    \includegraphics[width = 0.8\textwidth]{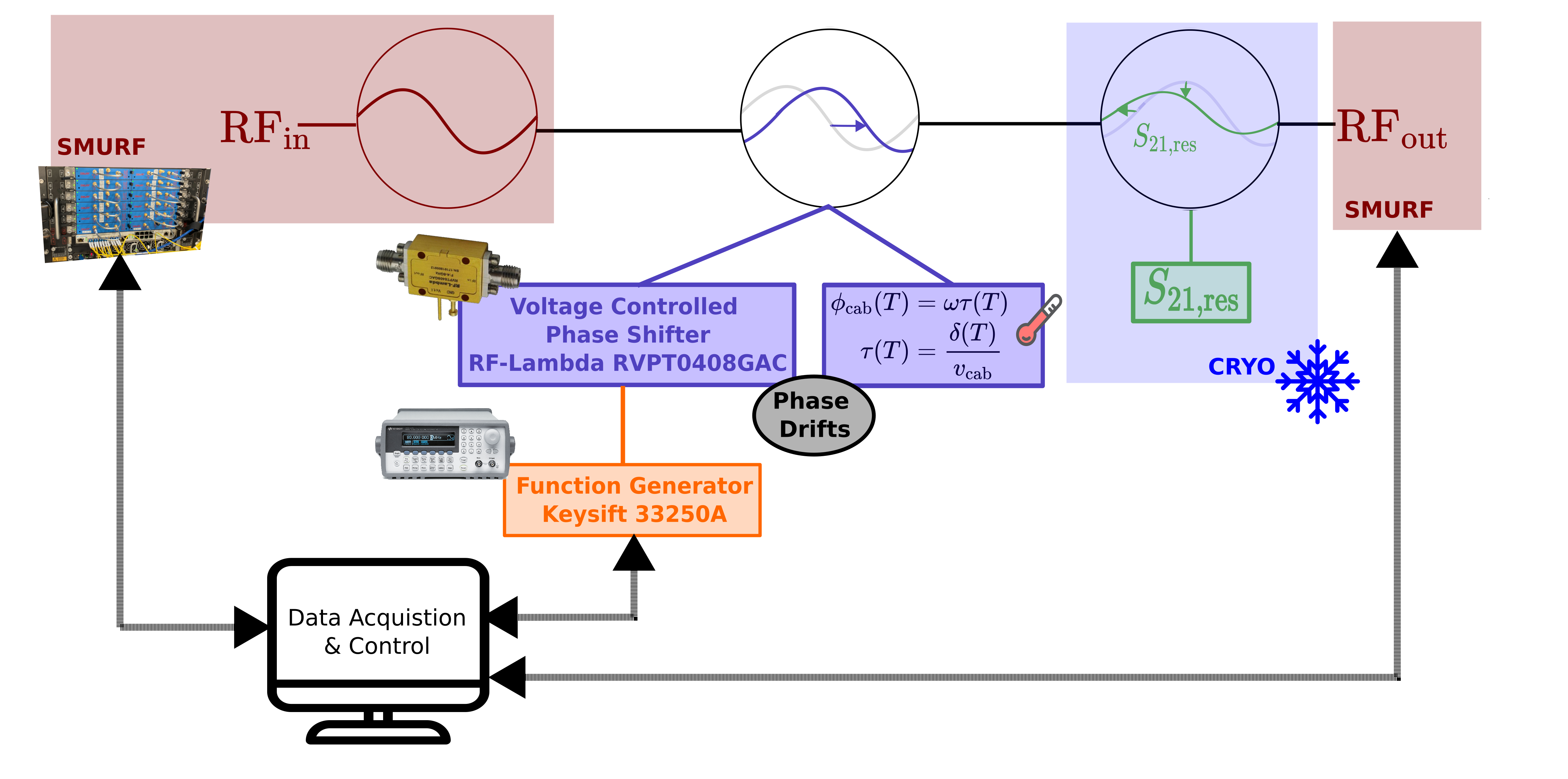}
    \caption{Measurement setup for demonstrating streamed data cleaning using pilot tones. The pilot tones and resonator probe tones are generated in the SMuRF (red box left), they then pass through a phase shifter (bold blue text) controlled by a function generator (bold orange text) set to a 30 deg phase peak-to-peak 10 mHz sine wave, then sent through the cryo cabling and resonators (blue box), and read back in with the SMuRF (red). All lab equipment is controlled with a common server and the data for both the resonator channels and pilot tones are co-sampled and written to disk at 200 Hz.}
    \label{fig:measurement_setup}
\end{figure}

A simplified measurement setup is shown schematically in figure \ref{fig:measurement_setup}. Shown in blue in the center is the round-trip phase drift present in our readout system. We allow the temperature of the room to drive the cable phase drift to test our assumptions about the scale of the temperature coupling to the cabling as discussed in section \ref{sec:measurement}. We alternatively use an analog phase shifter to inject a controlled phase signal to explore our cleaning method over different amplitude and timescales of phase injection as discussed in section \ref{sec:phase_injection}.

\subsection{Resonator Tuning}\label{sec:tuning}
\begin{figure}[h]
    \centering
    \includegraphics[width = 0.9\textwidth]{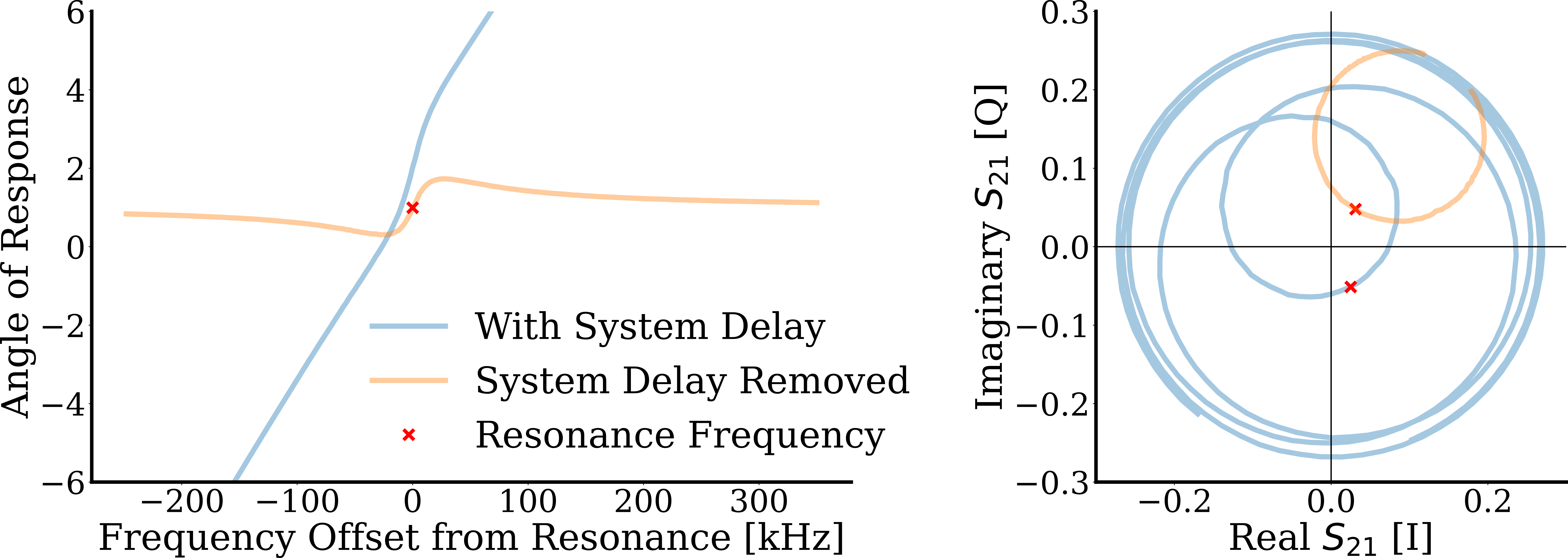}
    \vspace{0.25cm}
    \caption{The first step in the resonator tuning removes the fixed time delay in the system. The fixed time delay produces a large linear phase offset (left) as a function of frequency (blue curve); shown on the same scale as the corrected data (orange) (i.e. with $\sim$ 10 radians subtracted). If the fixed delay is left uncorrected it produces a phase slope on the same order as the slope through resonance, which complicates the resonance identification. When the delay is corrected (orange curve) the maximum phase slope is at resonance and the points far from resonance asymptote to the same value---in this case $\sim$1 radian. In the Real-Imaginary plane the fixed delay looks like many large loops (amplitude set by the loss around the resonance frequency) with a small distorted elliptical circle at the resonance. The resonance shape is both rotated and distorted (blue curve) relative to the corrected case (orange).}
    \label{fig:estimate_phase_delay}
\end{figure}

Our standard resonator tuning procedure is implemented in the \texttt{pysmurf}\footnote{\texttt{pysmurf} public Github repository: \href{https://github.com/slaclab/pysmurf}{https://github.com/slaclab/pysmurf}} user control software used to interface with the SMuRF at a high level. It consists of four main steps: 1) estimating overall phase delay, 2) finding resonances, and 3) estimating per-channel calibration. The phase delay estimation calculates the total time delay from the tone synthesis through to the tone channelization by sweeping a tone across a central window and fitting the phase versus frequency. The algorithm then adjusts delay parameters in the ADCs and channelized baseband processors to compensate for most of the $\sim$5-10 $\mu$S delay (dominated by the digital delays in firmware). If left uncalibrated, the phase slope from the cable delay alone is on the same order as the phase slope around resonance, as shown in the blue curve in figure \ref{fig:estimate_phase_delay}. Once properly compensated we can recover the orange curve, where the tails away from resonance asymptote to the same phase value around 1 radian and the phase slope is largest on resonance.

Next we coarsely sweep a tone across the full SMuRF bandwidth (4-6 GHz) with a step size of $\sim$40 kHz measuring the complex transmission at each point. Locations where there is concurrently a maximum in the transmission phase derivative and a minimum in the transmission magnitude are identified as candidate resonators. The gold curve in figure \ref{fig:pilot_tone_locs} is an example of coarse sweep data. This is followed by a series of fine sweeps with step size $\sim$ 2 kHz in 600 kHz windows centered around the frequencies identified in the coarse sweep. The orange data in figure \ref{fig:estimate_phase_delay} is an example of fine sweep data.

\begin{figure}[h]
    \centering
    \includegraphics[width = \textwidth]{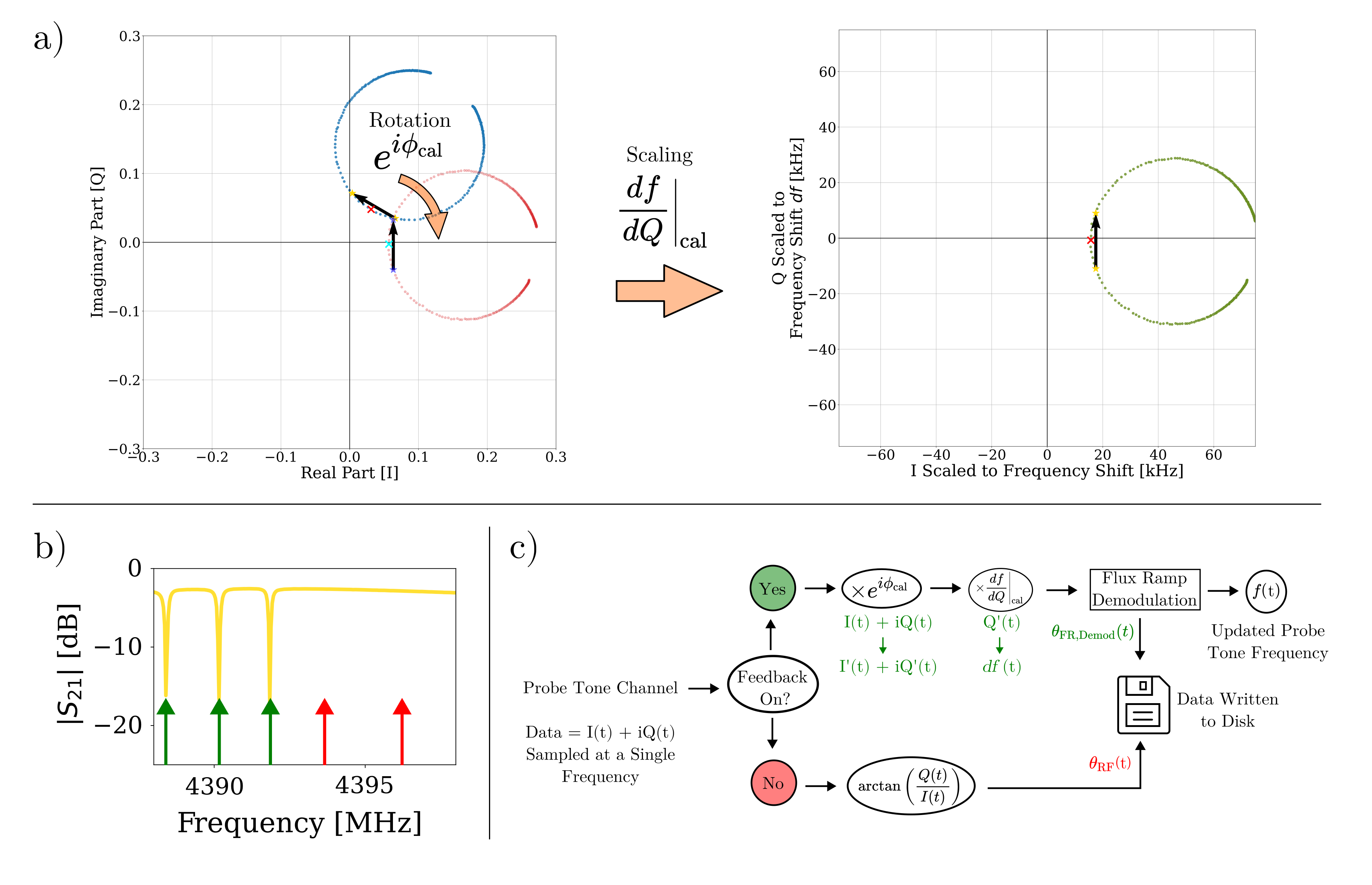}
    \caption{Description of probe tone processing. a) A graphical description of calibration steps applied to each feedback enabled probe tone channel, which include a rotation and scaling to convert from Q to resonance frequency shift. The black vector shown on all circles is used to calculate these calibration transformations. b) Transmission magnitude plot zoomed in around a location  with 3 resonances followed by a flat background transmission showing the distinct frequency locations where we place feedback enabled (green) and feedback disabled (red) tones. c) A flow chart showing the different processing steps between feedback enabled and feedback disabled tones. If feedback is enabled the complex data from the tone is rotated and scaled by the calibration factors to extract an estimate of resonance frequency shift, which is then fed to the flux ramp demodulation algorithm to update the tone frequencies and stream demodulated flux ramp phase data to disk. If feedback is disabled the complex data is converted to phase and this RF phase is then down-sampled and written to disk co-sampled with the feedback enabled channels.}
    \label{fig:tuning}
\end{figure}

This fine sweep is used to calculate a further per-channel calibration parameter which rotates the probe tone response such that the direction of maximum frequency response is aligned with the imaginary measurement axis and the measured signal in the imaginary axis is scaled into units of frequency shift. This calibration is shown graphically in figure \ref{fig:tuning}a. The angle of the vector relative to the imaginary axis determines the rotation calibration which transforms the blue curve into the red, aligning the Q axis with the direction of frequency shifts. The magnitude of the vector in units of Q voltage divided by the frequency offset between the two points used to calculate the vector provides the scaling from Q voltage to frequency shift, producing the green curve where the axes are now scaled to frequency shift from resonance. 

Figure \ref{fig:tuning}c describes the nominal data streaming when all tuning steps are complete. The data measured from a probe tone has two modes: 1) if a probe tone has feedback enabled, all of the aforementioned calibrations are applied to estimate the frequency shift, which is then fed to the flux ramp demodulation algorithm that outputs the demodulated flux ramp phase and an updated estimate of the resonance frequency. 2) if the feedback is disabled for a channel, then the calibration steps are skipped and RF phase is calculated directly, filtered, and down-sampled to the same data rate as the feedback-enabled channels and written to disk. 

In the current operational scheme all calibrations discussed in this section are assumed to be fixed over an observation. However, we know that this assumption can break down. In the rest of this article we describe one way the calibration can vary, uncompensated time delays. We estimate and measure their expected magnitude, and demonstrate a method for monitoring and cleaning contamination from these delays. The feedback off data-taking mode described in this section is a key firmware feature enabling phase drift monitoring and cleaning.

\section{Effects of Time Delay on the Resonator Transfer Function}\label{sec:res_tf}
\begin{figure}[h]
    \centering
    \includegraphics[width = 0.9\textwidth]{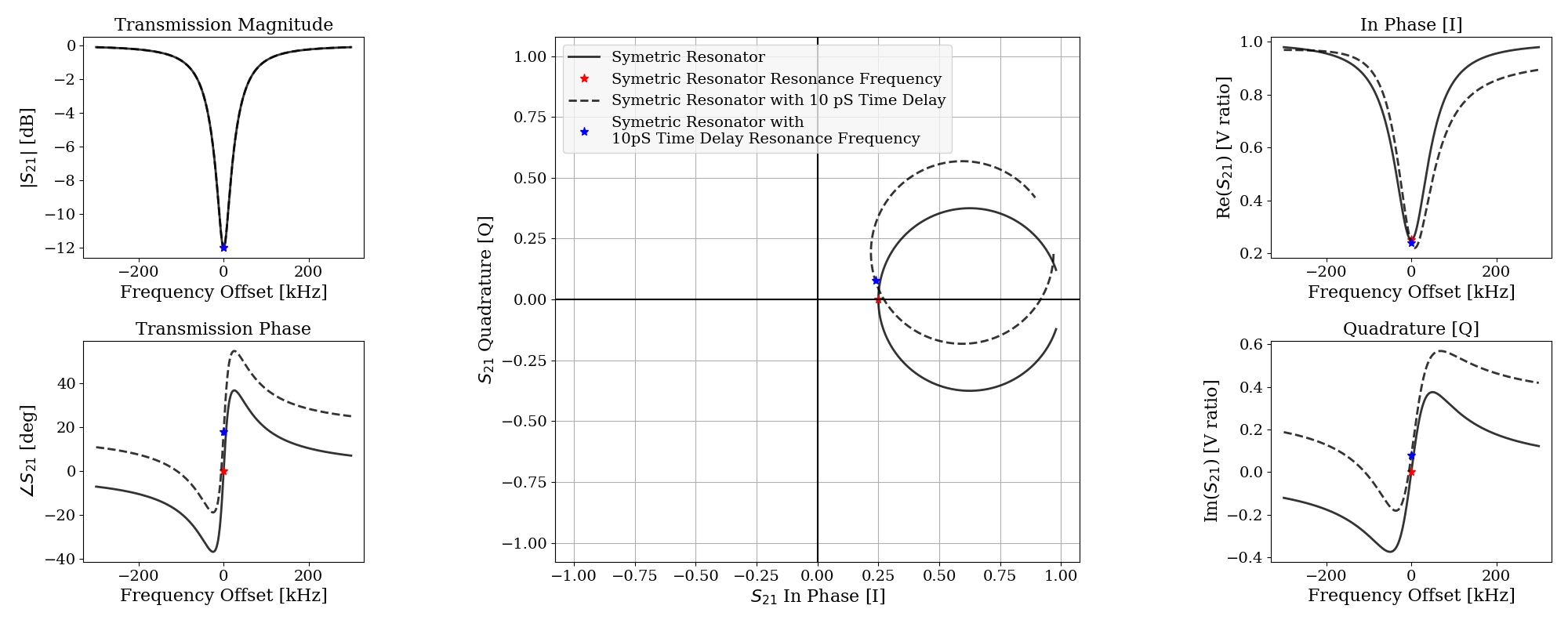}
    \caption{Resonator transmission $S_{21}$ at 0 cable delay (solid) and 10 pS delay (dashed) displayed in magnitude (upper left), phase (lower left), I-Q plane (center), real (upper right), and imaginary (lower right).}
    \label{fig:sym_res}
\end{figure}

Here we describe how a small uncalibrated time delay impacts the resonator transmission. A perfect quarter wave resonance with no time delay, asymmetry, or loss readout in transmission is described by equation \ref{eq:perfect_res}
\begin{equation}\label{eq:perfect_res}
    S_{21,\mathrm{res}}(x) = 1 - \frac{Q_r}{Q_c}\frac{1}{1+i2Q_rx}
\end{equation}
where $S_{21,\mathrm{res}}$ is the scattering matrix forward transmission parameter of the resonator without cabling delay, loss, etc. $Q_r$ is the resonator total quality factor, $Q_c$ is the coupling quality factor and is related to $Q_r$ and internal quality factor $Q_i$ by the relation $Q_r^{-1} = Q_c^{-1} + Q_i^{-1}$. $f$ is the frequency of the probe tone and $f_r$ is the resonance frequency. In the I-Q (real-imaginary) transmission plane this traces out a circle which crosses the I-axis at the resonance frequency. The vector tangent to the circle at resonance is parallel to the Q-axis and perpendicular to the I-axis. Figure \ref{fig:sym_res} shows the IQ circle along with the transmission in magnitude, phase, I, and Q vs frequency of the ideal resonator symmetric resonance.

The time delay $\tau$ is the physical time it takes for a guided EM wave to travel some distance $\ell$ in a waveguide. $\tau$ is set by the wavespeed $v$, which varies depending on the dielectric medium and geometry in the waveguide. Given a time delay the amount of phase delay is linearly dependent on frequency as given by equation \ref{eq:phase_from_tau}. 
\begin{equation}\label{eq:phase_from_tau}
    \theta = 2\pi\tau f
\end{equation}
The effect on the resonator transmission is a rotation of the resonator IQ circle and modifies equation \ref{eq:perfect_res} to:
\begin{equation}\label{eq:rotated_res}
    S_{21,\mathrm{meas}}(f) = e^{i2\pi\tau f}\left(1 - \frac{Q_r}{Q_c}\frac{1}{1+i2Q_r\left(\frac{f-f_r}{f_r}\right)}\right) = e^{i2\pi\tau f}S_{21,\mathrm{res}}
\end{equation}

\begin{figure}[h]
    \centering
    \includegraphics[width = 0.8\textwidth]{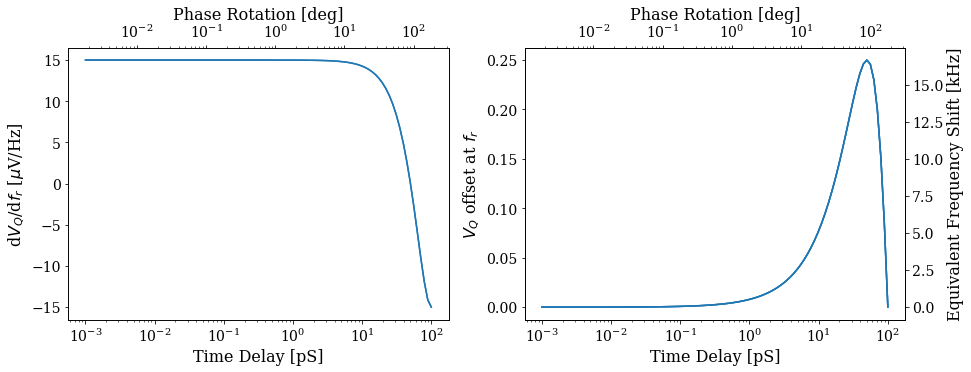}
    \caption{An uncompensated time delay produces two main effects: a mis-calibration of the $d$Q $\longrightarrow$ $df$ conversion, and a probe tone frequency shift. Change in imaginary transmission slope (left) characterizes the gain mis-calibration and projected Q component (right) characterizes the probe tone frequency drift shown on the right axis by converting the left axis with the $d$Q $\longrightarrow$ $df$ at $\tau$ = 0. Both effects are plotted for time delays between 1 fS and 100 pS.}
    \label{fig:dVqdf_vs_Tau}
\end{figure}

The rotation is visualized in figure \ref{fig:sym_res}, where a resonator with characteristic frequency of 5 GHz and 0 time delay (solid line) is compared with an equivalent resonator with 10 pS of time delay added (dashed line). There are two main impacts of the rotation when projected into the Q-axis: 1) the voltage measured in Q is no longer 0 when the probe tone is tuned to the resonance frequency (shown by a blue star in figure \ref{fig:sym_res}) and 2) the slope $dV_{\mathrm{Q}}/df$ evaluated at the resonance frequency, or conversion factor from voltage measured in Q to equivalent frequency shift, is not equal to the value at $\tau=0$. Figure \ref{fig:dVqdf_vs_Tau} shows the amount of slope change and $V_Q$ offset as a function of added time delay for a resonator at 5 GHz with axes shown in Q-axis units (left axis) as well as referred to effective frequency shift (right) axis converted via the $\left.\frac{df}{dQ}\right\rvert_{\mathrm{cal}}$ calibrated at $\tau$ = 0. Note that the up to roughly 10 degrees the slope is roughly constant and up to about 1 degree the Q offset is roughly constant. 

\section{Estimated Level of Time Delay Drift}\label{sec:expected_mag}
\begin{figure}[h]
    \centering
    \includegraphics[width = 0.8\textwidth]{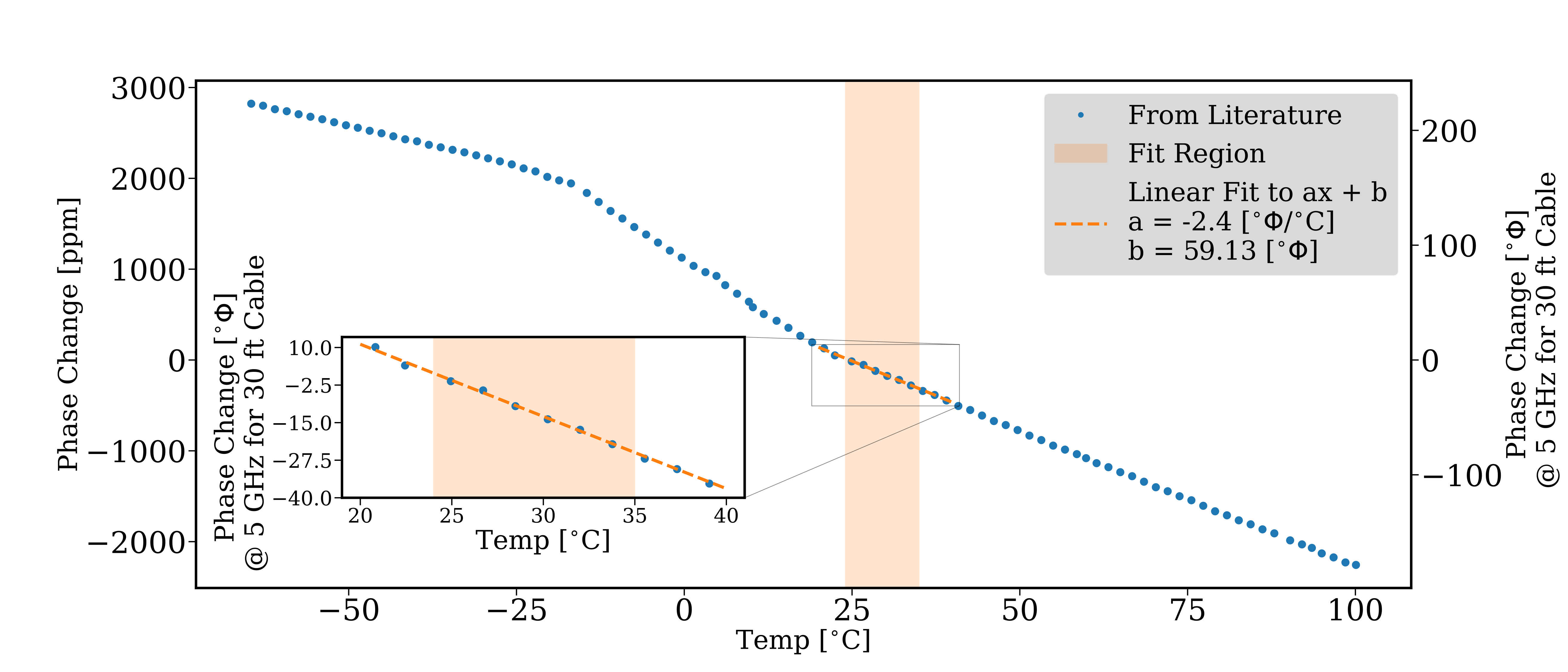}
    \caption{Fractional cable length change as a function of temperature converted to expected phase delay at 5 GHz for a 30 ft cable length. Data for Solid PTFE digitized from a Microwave Journal technical note\cite{PTFE_Compare} on differing PTFE temperature coefficients.}
    \label{fig:solid_ptfe}
\end{figure}
We assume (and confirm in section \ref{sec:measurement}) that the dominant time delay drift comes from the room temperature coaxial cables that connect between the cryostat vacuum jacket and the SMuRF electronics. Here we estimate the expected level of temperature drift given our cables and the thermal environment in our lab.

The dominant effect driving temperature drifts in the cables comes from expansion and contraction of the cable length driven mostly by the dielectric material. Most commonly used coaxial cables are constructed using a solid Polytetrafluoroethylene (PTFE) dielectric core, although there are other coaxial cables that use different dielectrics or lower temperature coefficient PTFE such as PTFE tape wrap, expanded PTFE\cite{Gore_ePTFE}, or Low Density PTFE foams. These have lower temperature coefficients, so an estimate of the induced phase from solid PTFE represents an upper bound. The main effect is a fractional change in length of the cable. The fractional change in length for solid PTFE can be found in the cable manufacturer literature\cite{PTFE_Compare} and is characterized by a knee around 25$^{\circ}$C. We estimate the contraction in the operating temperature range between 25$^{\circ}$ and 35$^{\circ}$C which is a linear range matched to the typical operating temperature range in our lab environment, as given in the inset of figure \ref{fig:solid_ptfe}. 
\begin{equation} \label{eq:ppm_to_deg}
    \delta\theta = 2\pi f\frac{\mathrm{ppm}}{10^6}\frac{\sqrt{\epsilon_r}}{c}\ell
\end{equation}
Equation \ref{eq:ppm_to_deg} is used to convert fractional length change in parts-per-mllion (ppm) to an expect phase change for a given cable length and probe tone frequency. Taking equation \ref{eq:ppm_to_deg} and the data shown in figure \ref{fig:solid_ptfe} we estimate 2.4 degrees phase per degree of temperature change at a 5 GHz readout frequency assuming that 30 ft of cable (15 ft. in and 15 ft. out) is used to connect between the cryostat and the SMuRF electronics, matching our test setup.
\begin{figure}[h]
    \centering
    \includegraphics[width = \textwidth]{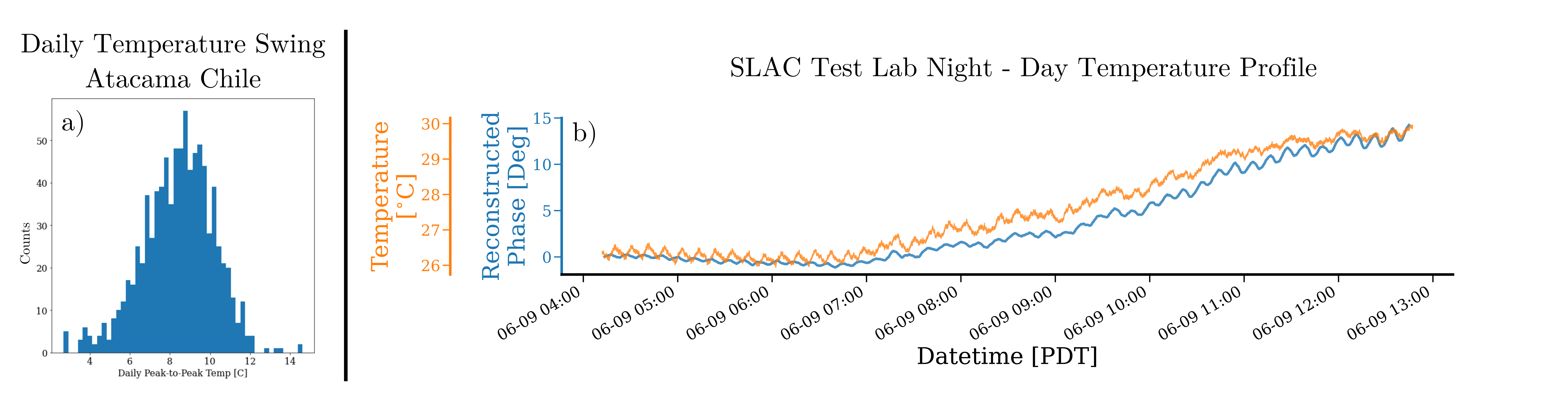}
    \caption{a) Histogram of the daily temperature swing at the APEX observation site in the the Atacama Desert in Chile between 2012-2017, during which a median daily swing is $\sim$10$^{\circ}$C. The Atacama Desert is a popular observing site for mm/sub-mm experiments which are particularly concerned with low frequency noise in their data. b) The typical night-to-day temperature variation in the test lab at the SLAC National Accelerator in Menlo Park, CA, USA (orange) and the corresponding phase shift at a resonator frequency as reconstructed from pilot tones (blue).}
    \label{fig:site_temp}
\end{figure}

Figure \ref{fig:site_temp}a shows a distribution of daily temperature swing over 5 years between 2012-2017 at the APEX observation site in the the Atacama Desert in Chile, a common observing site for mm/sub-mm and cosmic microwave background experiments which are particularly concerned with control of low-frequency noise in their data. Figure \ref{fig:site_temp}b shows a typical night-day temperature drift of $\sim4^{\circ}$C in our test lab at SLAC in June. The median daily temperature swing in Atacama is $\sim$ 8$^{\circ}$C, which means we expect an average day to induce up to 20 degrees of phase shift. Using shorter cables reduces this effect linearly and using alternative cable constructions with lower temperature coefficients provides some hardware mitigation. For some applications hardware mitigation may be sufficient but here we focus on a software correction method assuming the hardware mitigation cannot reduce this effect enough for our application.

\section{Thermal Time Delay Reconstruction}\label{sec:measurement} 
Here we set up a measurement to confirm: 1) the assumption that thermal drifts dominate our cable phase shift and 2) that the order of magnitude of the phase drift matches our expectations given cable construction and operating environment as discussed in section \ref{sec:expected_mag}.

\begin{figure}[h]
    \centering
    \includegraphics[width = \textwidth]{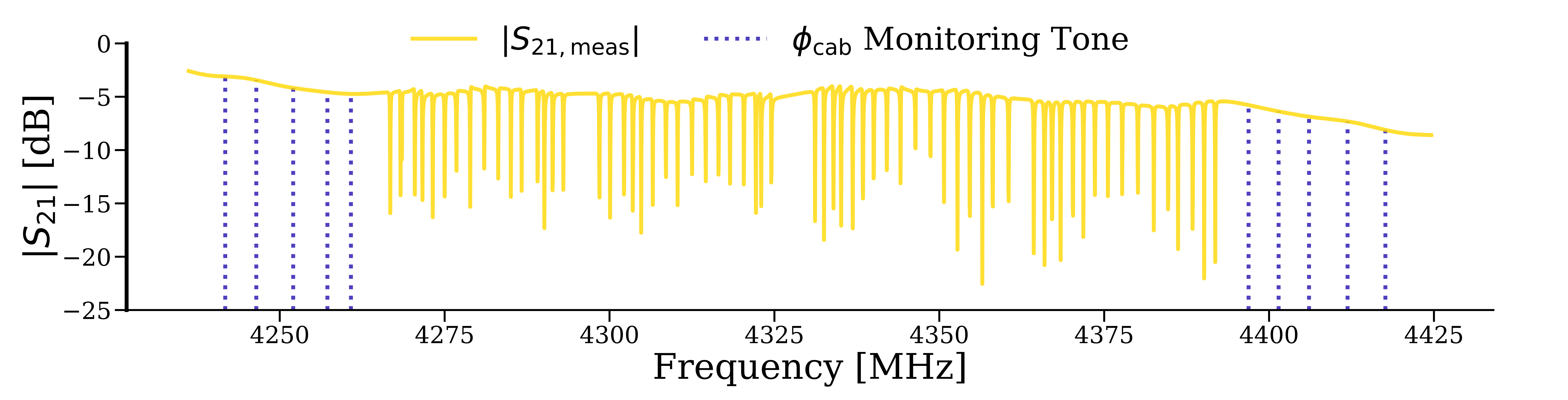}
    \caption{Transmission profile of a 64 channel resonator chip (solid gold) with the locations of the pilot tones in the 4-4.5 GHz band used for monitoring the variation in cable delay (dashed purple). Twenty additional tones are used between 4.5-5.5 GHz as well.}
    \label{fig:pilot_tone_locs}
\end{figure}

To measure the cable time delay we turn on a number of additional probe tones, called pilot tones, tuned to frequencies where there are no resonances and set to feedback off discussed in section \ref{sec:tuning}. The locations of the pilot tones used for this demonstration are shown in figure \ref{fig:pilot_tone_locs}. There were also 20 tones placed between 4.5-5.5 GHz. The yellow curve shows the $S_{21}$ with the resonance dips of the 64-channel multiplexer chip. Since the phase is linearly related to the time delay (equation \ref{eq:phase_from_tau}) we can estimate the time delay with a linear fit to the phase measured from the pilot tones versus their tone frequencies. We then use these fit results to calculate the phase delay at the tracked resonator frequencies.

We sample the phase from the pilot tones at 5 Hz. Once every 30 seconds we take the difference in phase $d\theta$ between the beginning and end of this interval and calculate the fit to $d\theta$ vs $f$. From the fit we calculate the phase changes at the resonance frequencies. This calculated phase change is shown in blue on the right panel of figure \ref{fig:site_temp}. We place a thermometer on \textit{one end} of the RF cables (orange) and show that the reconstructed phase shift qualitatively tracks the cable temperature, validating our assumption that the temperature coupling is dominating our phase drift. In the future we plan to implement an active feedback in which the calculated angle shifts are added to the resonator probe tone calibration angles on each feedback interval (in this example 30 seconds).

To maximize the magnitude of the effect we stream data for 6-8 hours between the daily low temperature and daily high temperature in our test lab. A plot of the temperature change over this period is shown in orange in the right panel of figure \ref{fig:site_temp}. There is a total temperature change of $\sim$ 4 $^{\circ}$C as well as a $\sim$10.5 minute (1.6 mHz) temperature oscillation of roughly 0.4$^{\circ}$C caused by the chiller discussed in section \ref{sec:hw_setup}.

\begin{figure}[h]
    \centering
    \includegraphics[width = \textwidth]{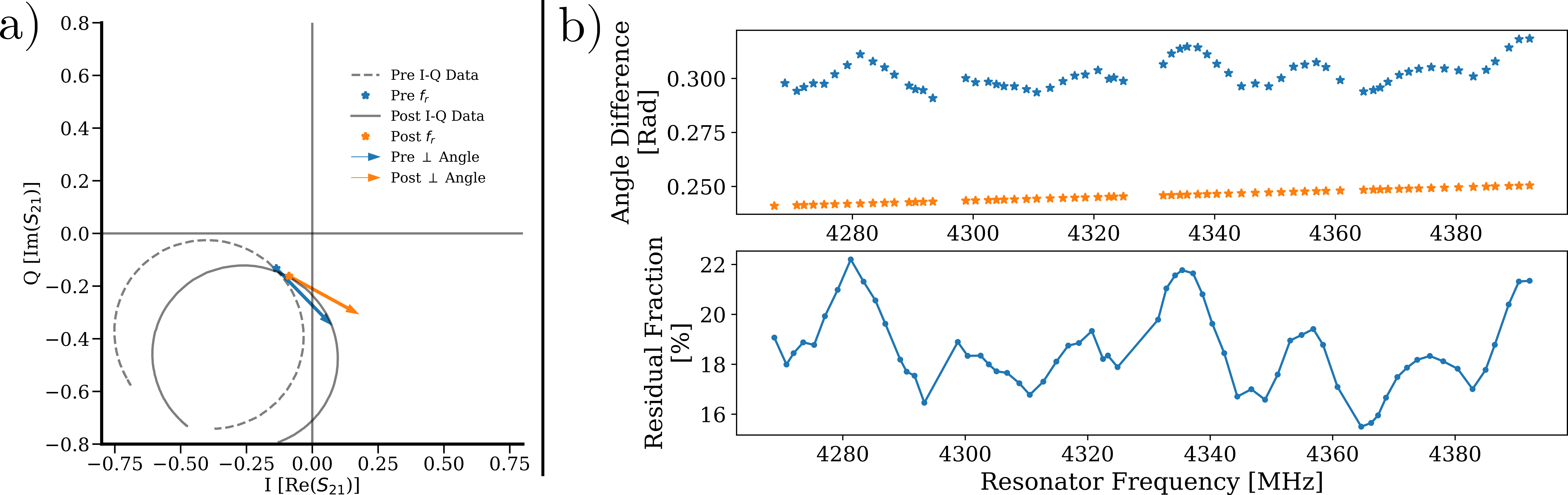}
    \caption{a) A single resonator IQ circle measured via sweeping the probe tone in frequency across the resonance before (dashed) and after (solid) the 7 hour night-to-day acquisition. The difference in the tangent to the circle (arrows) at the resonance frequency (stars) is used to estimate the amount of phase shift driven by the cable over the 7 hour acquisition. b) The estimate from the IQ circle (blue stars) is compared to the angle drift calculated from the fit to the pilot tones (orange stars) showing that applying feedback using the pilot tones would reduce the angle drift by $\geq$80\%. The oscillation in the blue data is suspected to be real and due to standing waves. The bias between the pilot tone and circle angle estimates is under investigation. }
    \label{fig:Compare_circl_to_pilot_tones}
\end{figure}

To confirm that the pilot tones are properly tracking the cable phase shifts we run the fine sweep discussed in section \ref{sec:tuning} at the beginning and end of the 7 hour dataset. We calculate the resonance circle rotation difference between the beginning and end by fitting the circle and calculating the tangent at the resonance frequency as shown in figure \ref{fig:Compare_circl_to_pilot_tones}a. We do this for all channels and compare with the offset angle calculated using the pilot tones during the stream as shown in figure \ref{fig:Compare_circl_to_pilot_tones}b. There is a repeatable systematic offset between the pilot tone estimate and the circle tangent estimate, where the pilot tones always return a smaller shift compared the $S_{21}$ tangent estimate method. Despite this, as shown in the bottom panel of figure \ref{fig:Compare_circl_to_pilot_tones}b we are still able to reduce the effect of the cable drift by $\geq$80\%. There is also an oscillation on top of the slope seen in the circle tangent angle estimate data (blue) but not the pilot tone angle estimate data (orange) due to the fact that we are only fitting the pilot tones to a linear model with frequency. However, if we turn on many more pilot tones across the same bandwidth such that we are densely mapping this oscillation period we do see this pattern in the difference data between two phases. We suspect that these oscillations are due to some standing waves in the system that are small, $\mathcal{O}$(5\%), compared with the absolute angle drift so that they represent a small correction to our angle estimate. Reducing this systematic offset is currently under investigation and will be presented in future work.

Using the natural environmental temperature change of the room to induce phase drifts we see that: 1) our estimate of the magnitude of thermal coupling of our cables was correct, 2) our assumption that the thermally induced phase drifts in the cable were dominating our phase drifts was correct and 3) that we could use the pilot tone method to track the phase drifts (up to the systematic offset and oscillations discussed above).

\section{Controlled Phase Injection}\label{sec:phase_injection}
After validating our assumptions about thermal coupling and our ability to monitor cable phase shown in section \ref{sec:measurement} we implemented an analog phase shifter to explore our ability to monitor and clean phase drifts over a controllable range of amplitudes and modulation rates. We used the RF-Lambda RVPT0408GAC analog phase shifter which produces a phase drift linearly proportional to the input voltage. To control the phase shifter we connected the analog voltage control pins to a Keysight 33250A function generator through a protection diode to prevent reverse biasing the phase shifter. The voltage controlled phase shifter measurement setup is shown in the left blue option for phase drift injection in figure \ref{fig:measurement_setup}.

\subsection{Cleaning Phase Drift Contamination}\label{sec:results}
To demonstrate our ability to use pilot tones to clean the contamination in the resonator readout channels due to the varying cable delay we input into a phase shifter a 10 mHz sine wave with a phase delay amplitude of $\sim$30 deg peak-to-peak and then acquire data with both the pilot tones and tracked resonator channels co-sampled at 200 Hz for 40 minutes. We then set the function generator sine wave modulation off and stream data for an additional 40 minutes to obtain a baseline. 

\begin{figure}[h]
    \centering
    \includegraphics[width = 0.8\textwidth]{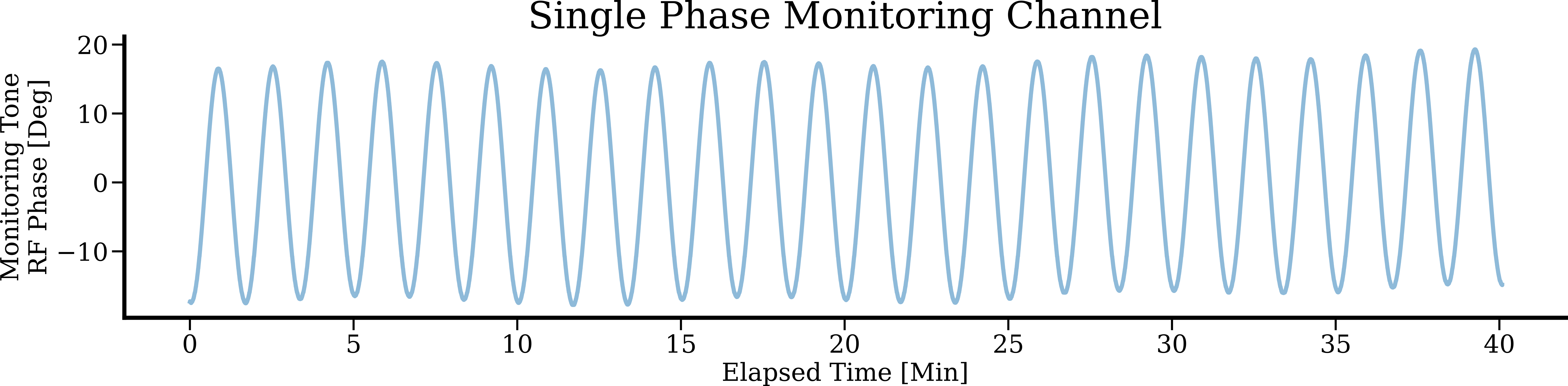}
    \\[\bigskipamount]
    \includegraphics[width = 0.8\textwidth]{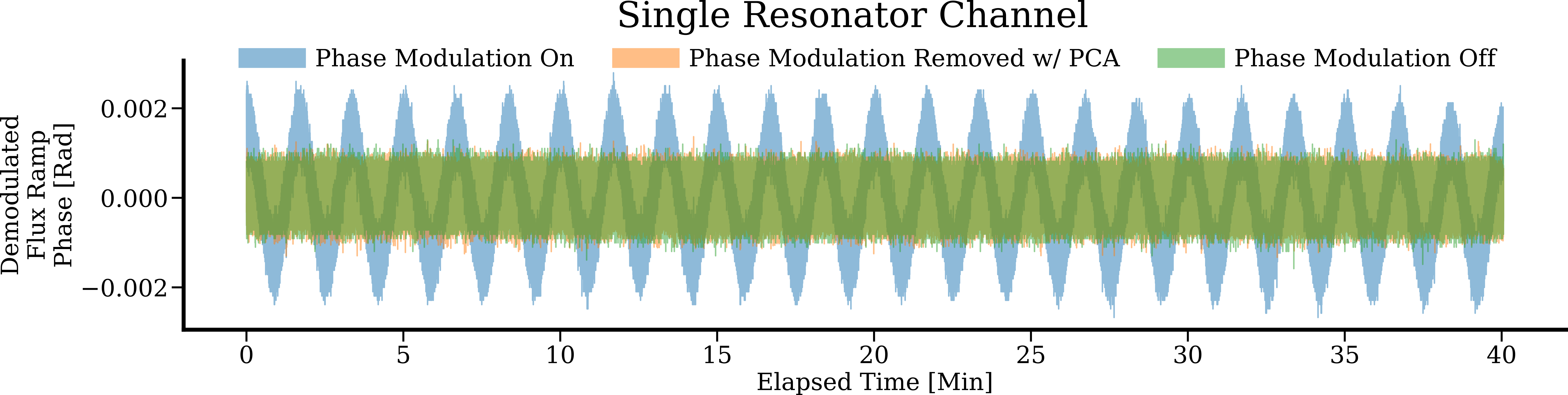}
    \\[\bigskipamount]
    \includegraphics[width = 0.9\textwidth]{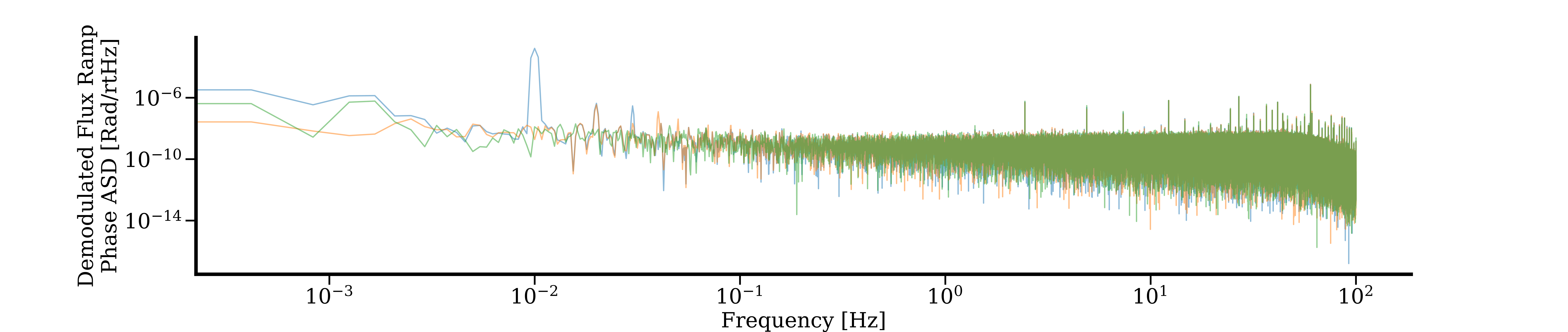}
    \caption{(Top) Single pilot tone channel with feedback off showing that it is following the 10 mHz phase modulation with a high significance, a single resonator time stream (center) and amplitude spectral density (bottom) for 40 minutes with the phase modulation on (blue) and off (orange) and with the phase modulation contamination cleaned from the blue curve using singular value decomposition (green).}
    \label{fig:results}
\end{figure}

The cleaning is implemented here with an SVD algorithm. We take the data sized $N\times T$ where $N$ is the number of channels and includes both the tracked resonator channels as well as the pilot tone channels and $T$ is the number of time samples and perform a singular value decomposition\cite{SVD}:
\begin{equation}
    \bold{S}_{N\times T} = \bold{W}_{N\times K}\bold{M}_{K\times T}
\end{equation}
Here $\bold{W}$ is the weights matrix size $N\times K$ where $K$ is the number of eigenmodes of the signal covariance matrix and $\bold{M}$ is the modes matrix. To get $\bold{W}$ and $\bold{M}$ we must first solve for the eigenvalues of the covariance matrix of the signal:
\begin{equation}
    \mathrm{Cov}[\bold{S}] = \bold{W}\bold{E}\bold{W}^{\mathrm{T}}
\end{equation}
We get the weights $\bold{W}$ directly as the eigenvectors and construct the modes $\bold{M} = \bold{W}^{\mathrm{T}} \bold{S}$. The modes represent a correlated signal in the time ordered data seen between many channels, while the weights scale that signal for each channel. The modes are sorted in order of strength. In our case, since all channels see the phase signal with correlation near 100\% across the pilot tone channels and the amplitude of that signal is much larger than the resonator-demodulated phase signal, the principle component (or first mode) that is picked out is the signal correlated with the phase modulation. Alternatively, instead of just picking out the first mode we could choose the mode (or combination of modes) that minimizes the pilot tone channel signal only and use that mode for cleaning. In our lab testing with resonators uncoupled to detectors there are very few other correlated noise sources, so we always found the principle component to be dominated by the pilot tone signal. We can then remove this mode from the resonator channels to clean the contamination as shown in figure \ref{fig:results}.

The green curve in the center and bottom plots of figure \ref{fig:results} compared the the orange curve shows the quality of our data cleaning. The time ordered data look nearly identical; however, one can see in the amplitude spectral density (ASD) that the odd harmonics of the 10 mHz injected signal appear to be cleaned better. There is also slightly different amounts of power in the lowest (poorly sampled) bins of the ASD, likely due to the SVD also cleaning some of the low frequency phase drift induced by the temperature drift in the room (and 1.6 mHz chiller cycle) discussed in section \ref{sec:measurement}. This demonstrates cleaning at a single phase injection amplitude and frequency however, for future work we plan to expand this to a range of frequencies and amplitudes to investigate over what parameter space this method is effective.

\section{Conclusion}\label{sec:conclusion}
We show that the phase drifts in our microwave SQUID multiplexer optimized for CMB applications are dominated by ambient temperature drifts coupling to our room temperature coaxial cables and that we can effectively monitor these phase drifts using a set of pilot tones set to frequencies where there are no resonances. Follow-up studies are required to identify and minimize the systematic offset between the angle shift calculated from the pilot tones and derived from the tangent vector to the $S_{21}$ resonance circle. However, we show contamination of the resonator channels from the cable delay variation and a post processing cleaning that largely removes this contamination using an SVD.

This is an important demonstration for next generation experiments that require extremely tight control of low-frequency correlated noise such as B-mode CMB surveys. Many resonator readout techniques could benefit from pilot tones because they offer a very clean measurement of the system phase drift independent of the resonator signal without adding any system noise. In the future we plan to use these monitoring tones to implement an active feedback in which the probe tone calibration phase is periodically adjusted over a timestream based on the calculated phase drifts from the pilot tones. Additionally, we plan to explore the data cleaning method presented here over a wide range of amplitudes and frequencies to evaluate the method for use in astronomical observatories such as cosmological B-mode searches, where control of low frequency noise and signal injection around particular modulator frequencies (i.e. half-wave plate or variable phase modulator) are particularly important for the final science data products.

\acknowledgments 
MSF was supported in part by the Department of Energy Office of Science Graduate Student Research (SCGSR) Program. The SCGSR program is administered by the Oak Ridge Institute for Science and Education (ORISE) for the DOE, which is managed by ORAU under contract number DE-SC0014664.

\bibliography{report} 
\bibliographystyle{spiebib} 

\end{document}